\documentclass{ws-ijmpa}

\begin{document}

\markboth{Yue-Liang Wu} {SO(3) Gauge Symmetry and Nearly
Tri-bimaximal Neutrino Mixing}

\catchline{}{}{}{}{}
%
%%%%%%%%%%%%%%%%%%%%%%%%%%%%%%%%%%%%%%%%%%%%%%%%%%%%%%%%%%%%%%%%%%%%

\title{SO(3) Gauge Symmetry and Nearly Tri-bimaximal Neutrino Mixing }

\author{Yue-Liang Wu}

\address{Kavli Institute for Theoretical Physics China, Institute of
Theoretical Physics \\ Chinese Academy of sciences, Beijing 100190,
China \\
ylwu@itp.ac.cn}

%\author{SECOND AUTHOR}

%\address{Group, Laboratory, Address\\
%City, State ZIP/Zone, Country\\
%second\_author@domain\_name}

\maketitle

\begin{history}
\received{22 June 2008}
%\revised{Day Month Year}
\end{history}

\begin{abstract}
In this note I mainly focus on the neutrino physics part in my talk
and report the most recent progress made in \cite{YLW0}. It is seen
that the Majorana features of neutrinos and SO(3) gauge flavor
symmetry can simultaneously explain the smallness of neutrino masses
and nearly tri-bimaximal neutrino mixing when combining together
with the mechanism of approximate global U(1) family symmetry. The
mixing angle $\theta_{13}$ and CP-violating phase are in general
nonzero and testable experimentally at the allowed sensitivity. The
model also predicts the existence of vector-like Majorana neutrinos
and charged leptons as well as new Higgs bosons, some of them can be
light and explored at the LHC and ILC.

\end{abstract}

\ccode{PACS numbers: 14.60.Pg, 12.60.-i, 11.30.Hv,14.60.St}

%\maketitle
\section*{}

The observed massive neutrinos and dark matter in our universe
challenge the standard models of both particle physics and
cosmology. So both flavor physics and cosmology may tell us
fundamental physics\cite{JN}. In this note I will concentrate on
discussing the neutrino physics.

Various neutrino
experiments\cite{EXP1,EXP2,EXP3,EXP4,EXP5,EXP6,EXP7,EXP8,PDG}
provide more and more stringent constraints on the three mixing
angles and mass squared differences for three
neutrinos\cite{SV,TNM2,TNM1}
\begin{eqnarray}
& & 30\mbox{}^{\circ}<\theta_{\mathrm{12}}<38\mbox{}^{\circ}, \quad
36\mbox{}^{\circ}<\theta_{\mathrm{23}}<54\mbox{}^{\circ}, \quad
\theta_{\mathrm{13}}<10\mbox{}^{\circ}  \\
& & 7.2\times{10}^{-5}\ {\mathrm{eV}}^{2}    <\Delta
m_{21}^{2}=m_{\nu_{\mu}}^2 - m_{\nu_e}^2 <8.9\times {10}^{-5}\ {\mathrm{eV}}^{2}, \nonumber \\
& & 2.1\times{10}^{-3}\ {\mathrm{eV}}^{2}    <\Delta m_{32}^{2}=
m_{\nu_{\tau}}^2 - m_{\nu_{\mu}}^2 <3.1\times {10}^{-3}\
{\mathrm{eV}}^{2}
\end{eqnarray}
at the 99\% confidence level\cite{SV}. In comparison with the quark
sector, it raises a puzzle that why neutrino masses are so tiny, but
their mixing angles are so large\cite{HF,PM}. The only peculiar
property for neutrinos is that they could be Majorana fermions, so a
natural solution to the puzzle is most likely attributed to the
Majorana features. Thus revealing the origin of large mixing angles
and small masses of neutrinos is important not only for
understanding neutrino physics, but also for exploring new physics
beyond the standard model\cite{MS}.

The current data on the neutrino mixing angles are consistent with
the so-called tri-bimaximal mixing with $\theta_{12} =
\sin^{-1}(1/\sqrt{3}) = 35^{\circ}$, $\theta_{23} =
\sin^{-1}(1/\sqrt{2}) = 45^{\circ}$ and $\theta_{13} =0$, which was
first proposed by Harrison, Perkins and Scott \cite{HPS}, and
investigated by various groups\cite{HPS1,HPS2,HPS3,HPS4}. The mixing
angle $\theta_{\mathrm{13}}$ is the most unclear parameter and
expected to be measured in near future. Such a tri-bimaximal mixing
matrix has been found to be yielded by considering some interesting
symmetries, especially the discrete symmetries\cite{Ma}. In general,
it is shown that the discrete symmetries lead to $\theta_{13} =0$
\cite{Lam}. In such a case, it is hard to be directly tested
experimentally. Alternatively, it is interesting to consider a
non-abelian gauge family symmetry
SO(3)\cite{YLW0,YLW1,YLW2,YLW3,YLW4,YLW5,YLW6,CS,MA,CW,BHKR} instead
of discrete symmetries discussed widely in literature. In this case,
the tri-bimaximal neutrino mixing matrix is generally obtained as
the lowest order approximation from diagonalizing a special
symmetric mass matrix\cite{YLW0}. In fact, the greatest success of
the standard model (SM) is the gauge symmetry structure
$SU(3)_{c}\times SU_{L}(2)\times U_{Y}(1)$ which has been tested by
more and more precise experiments. SO(3) gauge family symmetry can
be regarded as a simple extension of the standard model with three
families and Majorana neutrinos. It is noted that only SO(3) rather
than SU(3) is allowed due to the Majorana feature of neutrinos.

The $SO(3)\times SU(2)_{L}\times U(1)_{Y}$ invariant Lagrangian for
Yukawa interactions of leptons with Majorana neutrinos can be
constructed as follows\cite{YLW0}
\begin{eqnarray}
{\cal L}_Y & = & y_{\nu} \bar{l} \tilde{H}\nu_R + y_{N} \bar{l} H_N
N  + \frac{1}{2}\xi_{N} \bar{N} \Phi_{\nu} N  +
\frac{1}{2} M_R \bar{\nu}_R\nu_R^c  \nonumber \\
& + & y_e \bar{l} H E + \xi_e \phi_s \bar{E} e_R +  \frac{1}{2}
\xi_E \bar{E} \Phi_e E + H.c.
\end{eqnarray}
where $y_{\nu}$, $y_e$, $y_N$, $\xi_e$, $\xi_N$ and $\xi_E$ are all
real Yukawa coupling constants and $M_R$ is the mass of right-handed
Majorana neutrinos. All the fermions $\nu_{Li}$, $\nu_{Ri}$,
$e_{Li}$, $e_{Ri}$, $E_i$ and $N_{i}$ $(i=1,2,3)$ belong to SO(3)
triplets in family space. Where $\bar{l}_i= (\bar{\nu}_{Li},
\bar{e}_{Li} )$ denote $SU_L(2)$ doublet leptons, $H$ and $H_N$ are
$SU_L(2)$ doublet Higgs bosons with $\tilde{H} = \tau_2 H^*$.
$\nu_{Ri}$ are the right-handed neutrinos with $\nu_{Ri}^c =
c\bar{\nu}_{Ri}^T$ the charge conjugated ones. $E_i$ are $SU_L(2)$
singlet vector-like charged leptons and the $N_{i}$ are $SU_L(2)$
singlet vector-like Majorana neutrinos with $N_{ i}^c = N_{i}$.
$\phi_s$ is a singlet Higgs boson. The scalar fields $\Phi_{\nu}$
and $\Phi_e$ are SO(3) tri-triplets Higgs bosons satisfying
$\Phi_{\nu} = \Phi_{\nu}^{\ast}$, $\Phi_{\nu} = \Phi_{\nu}^T$ and
$\Phi_e = \Phi_e^{\dagger}$ which is required by the hermiticity
condition of Lagrangian and the Majorana condition of vector-like
neutrinos. The above Lagrangian is solely ensured by the following
discrete symmetry ($Z_2$ and $Z_4$)
\begin{eqnarray}
N \to i\gamma_5\ N, \quad \Phi_{\nu} \to - \Phi_{\nu}, \quad H_N \to
-iH_N, \quad \phi_s \to - \phi_s,\quad e_R \to - e_R
\end{eqnarray}
In terms of SO(3) representation, one can reexpress the real
symmetric tri-triplet Higgs boson into the following general form
\begin{eqnarray}
& & \Phi_{\nu}  \equiv  O_{\nu} \phi_{\nu} O_{\nu}^T, \quad
O_{\nu}(x) = e^{i\lambda^i \Theta^{\nu}_i(x)},  \quad \phi_{\nu}(x)
= \left(
                     \begin{array}{ccc}
                       \phi_1^{\nu} & \phi_2^{\nu} & \phi_3^{\nu} \\
                       \phi_2^{\nu} & \phi_3^{\nu} & \phi_1^{\nu} \\
                       \phi_3^{\nu} & \phi_1^{\nu} & \phi_2^{\nu} \\
                     \end{array}
                   \right)
\end{eqnarray}
with $\lambda^i$ $(i=1,2,3)$ being the generators of SO(3). Where
$\Theta_i^{\nu}(x)$ $(i=1,2,3)$ may be regarded as three rotational
scalar fields of SO(3), and $\phi_i^{\nu}(x)$ $(i=1,2,3)$ are three
dilatation scalar fields.

SO(3) gauge invariance allows us to fix the gauge by making SO(3)
gauge transformation $g(x)$, so that $g(x) \equiv O_{\nu}(x) \in
SO(3)$, we then arrive at the following Yukawa interactions
\begin{eqnarray}
{\cal L}_Y & = & y_{\nu} \bar{l} \tilde{H}\nu_R + y_{N} \bar{l} H_N
N + \frac{1}{2}\xi_{N} \bar{N} \phi_{\nu} N  +
\frac{1}{2} M_R \bar{\nu}_R\nu_R^c   \nonumber \\
& + & y_e \bar{l} H E + \xi_e \phi_s \bar{E} e_R +  \frac{1}{2}
\xi_E \bar{E} \hat{\Phi}_e E + H.c.
\end{eqnarray}
which is invariant under $Z_3$ transformation. Where $\hat{\Phi}_e =
O_{\nu}^T \Phi_e  O_{\nu}$ remains Hermitian and contains nine
independent scalar fields, which can generally be reexpressed in
terms of SO(3) representation as the following form
\begin{eqnarray}
 \hat{\Phi}_e \equiv U_e \phi_e U_e^{\dagger},
 \quad U_e(x) \equiv P_e O_{e}, \quad \quad O_e(x) =
e^{i\lambda^i \chi^{e}_i(x)},
\end{eqnarray}
and
\begin{eqnarray}
& & P_e(x)  =  \left(
 \begin{array}{ccc}
  e^{i\eta_1^e(x)} & 0 & 0 \\
    0 & e^{i\eta_2^e(x)} & 0 \\
    0 & 0 & e^{i\eta_3^e(x)}
  \end{array}
 \right),\quad \phi_{e}(x) = \left(
                     \begin{array}{ccc}
                       \phi_1^{e}(x) & 0 & 0 \\
                       0 & \phi_2^{e}(x) & 0 \\
                       0 & 0 & \phi_3^{e}(x) \\
                     \end{array}
                    \right)
\end{eqnarray}
where $\chi_i^e(x)$ $(i=1,2,3)$ are regarded as three rotational
scalar fields of SO(3), $\eta_i^e(x)$ $(i=1,2,3)$ denote three phase
scalar fields and $\phi_i^e(x)$ $(i=1,2,3)$ are three dilation
scalar fields.

We now consider the following general vacuum structure of scalar
fields under the above gauge fixing condition
\begin{eqnarray}
& & <H(x)> = v, \qquad <H_N(x)> = v_N  \nonumber \\
& &  <\phi_s(x)> = v_s, \qquad <\phi_i^{\nu}(x)> = v_i^{\nu},  \\
& &  <\phi_i^e(x)> = v_i^e,\qquad <\chi_i^e(x)> = \theta_i^e , \quad
<\eta_i^e(x)> = \delta_i^e \nonumber
\end{eqnarray}
namely $<P_e> \equiv P_{\delta}^e  = diag.( e^{i\delta_1^e}$,
$e^{i\delta_2^e}, e^{i\delta_3^e} )$ and $<O_e> = e^{i\lambda^i
\theta_i^e}$. Here $\delta_i^e$ (i=1,2,3) are CP phases arising from
spontaneous symmetry breaking and $\theta_i^e$ are three rotational
angles of SO(3).

Such a vacuum structure after spontaneous symmetry breaking leads to
the following mass matrices for neutrinos with a type II like
see-saw mechanism and for charged leptons with a generalized see-saw
mechanism
\begin{eqnarray}
& & M_{\nu} = m^D_{\nu} M_R^{-1} m^D_{\nu} + m^D_{N} M_N^{-1}
m^D_{N}, \\
& &  M_e =  V_e m_E^D M_E^{-1} m_E^D V_e^{\dagger}
\end{eqnarray}
with $m^D_{\nu} = y_{\nu} v$, $m^D_{N} = y_N v_N$, $m_E^D =
\sqrt{y_{e} \xi_e v  v_s }$, $V_e = <U_{e}> = P_{\delta}^e e^{i
\lambda^i \theta_i^e }$, and
\begin{eqnarray}
 & & M_N = \xi_{N} \left(
        \begin{array}{ccc}
          v_1^{\nu} &  v_2^{\nu} &  v_3^{\nu} \\
           v_2^{\nu} &  v_3^{\nu} &  v_1^{\nu} \\
           v_3^{\nu} &  v_1^{\nu} &  v_2^{\nu} \\
        \end{array}
      \right),\qquad  M_E = \xi_E \left(
            \begin{array}{ccc}
               v_1^{e} & 0 & 0 \\
              0 &  v_2^{e} & 0 \\
              0 & 0 &  v_3^{e} \\
            \end{array}
          \right) \\
%          ,\quad P_{\delta}^e = \left(
%             \begin{array}{ccc}
%               e^{i\delta_1^e} & 0 & 0 \\
%               0 &  e^{i\delta_2^e}  & 0 \\
%               0 & 0 &  e^{i\delta_3^e}  \\
%             \end{array}
%           \right) \\
& & V_e \equiv
               P_{\delta}^e \left(
                     \begin{array}{ccc}
                       c_{12}^ec_{13}^e\ \ & s_{12}^ec_{13}^e\ \ & s_{13}^e \\
                       -s_{12}^ec_{23}^e-c_{12}^es_{23}^es_{13}^e\  \ &
                       c_{12}^ec_{23}^e - s_{12}^es_{23}^es_{13}^e\ \ & s_{23}^e c_{13}^e \\
                       s_{12}^e s_{23}^e - c_{12}^ec_{23}^es_{13}^e\ \ & -c_{12}^es_{23}^e - s_{12}^ec_{23}^es_{13}^e\  \
                       & c_{23}^ec_{13}^e \\
                     \end{array}
                   \right)
\end{eqnarray}
with $c_{ij}^e \equiv \cos\theta_{ij}^e$ and $s_{ij}^e \equiv
\sin\theta_{ij}^e$, and $\theta_{ij}^e$ are given as functions of
$\theta_i^e$ $(i=1,2,3)$. A similar special symmetric neutrino mass
matrix like $M_N$ was also resulted for the Dirac-type neutrinos
with a new symmetry\cite{FLee} and the Majorana-type neutrinos with
the $Z_3$ group\cite{HWW}.

When taking the Majorana neutrino masses $M_R$ and $M_N$ to be
infinity large, the interactions with Majorana neutrinos decouple
from the theory,
\begin{eqnarray}
\frac{y^2_{\nu}}{M_R} \bar{l}\tilde{H} \tilde{H}^{T} l^c, \ \
\frac{y^2_{N}}{M_N} \bar{l}H_N H_N^{T} l^c \to 0, \quad \mbox{for}
\quad M_R,\   \  M_N \to \infty
\end{eqnarray}
which implies that the resulting Yukawa interactions in this limit
generate additional global U(1) family symmetries for the charged
lepton sector. Namely, once the Majorana neutrinos become very
heavy, the Yukawa interactions possess approximate global U(1)
family symmetries. Thus when applying the mechanism of approximate
global U(1) family symmetries\cite{HW,WW,WU} to the Yukawa
interactions after SO(3) symmetry is broken down spontaneously, we
have $M_R \gg m^D_{\nu}$, $M_N \gg m_N^D$ and $\theta_i^e \ll 1$,
namely
\begin{eqnarray}
 M_{\nu} \ll 1, \qquad \theta_{ij}^e \ll 1
\end{eqnarray}
which provides a possible explanation why the observed left-handed
neutrinos are so light and meanwhile the charged lepton mixing
angles must be small. Thus the neutrino mass matrix is given by a
type II like see-saw mechanism and the charged lepton mass matrix is
presented by a generalized see-saw mechanism. By diagonalizing the
mass matrices, we obtain
\begin{eqnarray}
V_{\nu}^{T} M_{\nu} V_{\nu} = diag.(m_{\nu_e}, m_{\nu_{\mu}},
m_{\nu_{\tau}} ), \quad V_e^{\dagger} M_e V_e = diag.(m_e, m_{\mu},
m_{\tau} )
\end{eqnarray}
where
\begin{eqnarray}
V_{\nu} = \left(
  \begin{array}{ccc}
    \frac{2}{\sqrt{6}}c_{\nu}\  \
     & \frac{1}{\sqrt{3}}\  \  & \frac{2}{\sqrt{6}}s_{\nu} \\
    -\frac{1}{\sqrt{6}}c_{\nu} - \frac{1}{\sqrt{2}}s_{\nu}\  \  & \frac{1}{\sqrt{3}}\  \  &
    \frac{1}{\sqrt{2}}c_{\nu} - \frac{1}{\sqrt{6}}s_{\nu} \\
    -\frac{1}{\sqrt{6}}c_{\nu} + \frac{1}{\sqrt{2}}s_{\nu}\  \  & \frac{1}{\sqrt{3}}\  \  &
    - \frac{1}{\sqrt{2}}c_{\nu}-\frac{1}{\sqrt{6}}s_{\nu}  \\
  \end{array}
\right) \equiv V_0 V_1
\end{eqnarray}
with
\begin{eqnarray}
 V_0 = \left(
  \begin{array}{ccc}
    \frac{2}{\sqrt{6}}
     & \frac{1}{\sqrt{3}} & 0 \\
    -\frac{1}{\sqrt{6}} & \frac{1}{\sqrt{3}} & \frac{1}{\sqrt{2}} \\
    -\frac{1}{\sqrt{6}} & \frac{1}{\sqrt{3}} & -\frac{1}{\sqrt{2}} \\
  \end{array}
\right), \quad V_1 = \left(
        \begin{array}{ccc}
          c_{\nu} & 0 & s_{\nu} \\
          0 & 1 & 0 \\
          -s_{\nu} & 0 & c_{\nu} \\
        \end{array}
      \right)
\end{eqnarray}
Here $V_0$ is the so-called tri-bimaximal mixing matrix\cite{HPS}.
For short, we have introduced the notations $c_{\nu} \equiv \cos
\theta_{\nu}$ and $s_{\nu} \equiv \sin \theta_{\nu}$ with
\begin{eqnarray}
\tan 2 \theta_{\nu} = \frac{\sqrt{3}(v_{21}^{\nu} - v_{31}^{\nu})
}{v_{21}^{\nu} + v_{31}^{\nu}}, \quad v_{21}^{\nu} \equiv
v_{2}^{\nu} - v_{1}^{\nu}, \quad v_{31}^{\nu} \equiv v_{3}^{\nu} -
v_{1}^{\nu}
\end{eqnarray}

As the smallness of charged lepton mixing angles $\theta_{ij}^e$
$(i<j)$ can be attributed to the mechanism of approximate global
U(1) family symmetries in the Yukawa sector, in a good approximation
up to the first order of $s_{ij}^e$, the leptonic CKM-type mixing
matrix in the mass eigenstates of leptons may be expressed as the
following simplified form
\begin{eqnarray}
V = V_e^{\dagger} V_{\nu} \simeq & \left(
                       \begin{array}{ccc}
                       1 \ & -s_{12}^e \
                        & - s_{13}^e \\
                       s_{12}^e \ &  1 \
                       & -s_{23}^e  \\
                       s_{13}^e \ &  s_{23}^e  \
                       & 1 \\
                     \end{array}
                   \right) P_{\delta}^{e \dagger}
           \left(
  \begin{array}{ccc}
    \frac{2}{\sqrt{6}}
     & \frac{1}{\sqrt{3}} & 0 \\
    -\frac{1}{\sqrt{6}} & \frac{1}{\sqrt{3}} & \frac{1}{\sqrt{2}} \\
    -\frac{1}{\sqrt{6}} & \frac{1}{\sqrt{3}} & -\frac{1}{\sqrt{2}} \\
  \end{array}
\right) \left(
        \begin{array}{ccc}
          c_{\nu} & 0 & s_{\nu} \\
          0 & 1 & 0 \\
          -s_{\nu} & 0 & c_{\nu} \\
        \end{array}
      \right)
\end{eqnarray}

The three vector-like heavy Majorana neutrino masses are obtained
via diagonalizing the mass matrix $M_N$, i.e., $V^{\dagger}_{\nu}
M_N V_{\nu} = diag.(m_{N_1}, \ m_{N_2},\ m_{N_3})$
\begin{eqnarray}
m_{N_1} = - m_N \sqrt{1-\Delta},\quad m_{N_2} = m_N, \quad m_{N_3} =
m_N \sqrt{1-\Delta}
\end{eqnarray}
with
\begin{eqnarray}
m_N \equiv \xi_N (v_1^{\nu} + v_2^{\nu} + v_3^{\nu} ), \quad \Delta
= 3(v_1^{\nu}v_2^{\nu} + v_2^{\nu}v_3^{\nu} +
v_3^{\nu}v_1^{\nu})/(v_1^{\nu} + v_2^{\nu} + v_3^{\nu})^2
\end{eqnarray}
The masses of three left-handed light Majorana neutrinos are given
in the physics basis as follows
\begin{eqnarray}
m_{\nu_e} = \bar{m}_{0} - m_1 ( 2 + \bar{\Delta}),\quad
m_{\nu_{\mu}} = \bar{m}_{0},\quad  m_{\nu_{\tau}} = \bar{m}_{0} +
m_1 \bar{\Delta}
\end{eqnarray}
with $\bar{m}_{0} \equiv m_0 + m_{1}$, $\bar{\Delta} =
1/\sqrt{1-\Delta} - 1$, $m_0 = (m^D_{\nu})^2/M_R$ and $ m_1 =
(m_N^D)^2/m_N$. It then enables us from the experimentally measured
neutrino mass squire differences to extract two mass parameters
$m_0$ and $m_1$ for a given value of parameter $\Delta$ with $\Delta
\neq 1$.

The numerical results are presented in table 1 by taking the central
values of the mass squire differences $\Delta m_{21}^2 =
m_{\nu_{\mu}}^2 - m_{\nu_{e}}^2 = 8 \times 10^{-5}\ eV^2$ and
$\Delta m_{32}^2 = m_{\nu_{\tau}}^2 - m_{\nu_{\mu}}^2 = 2.5 \times
10^{-3}\ eV^2$.
\begin{table}[ph]
\tbl{The masses of neutrino mass eigenstates $\nu_e$, $\nu_{\mu}$,
$\nu_{\tau}$ and mass parameters $m_0$ and $m_1$ as the functions of
the vacuum parameter $\Delta$.(The minus sign of the Majorana
neutrino mass can be absorbed by a redefinition of the neutrino
field $\nu_{e} \to i \nu_{e}$)} {\begin{tabular}{|c|c|c|c|c|c|}
  \hline
  $\Delta$ (input) & $m_0 (10^{-2} eV)$ & $m_1 (10^{-2} eV)$ & $m_{\nu_e} ( 10^{-2} eV)$ & $m_{\nu_{\mu}}(10^{-2} eV)$
  & $m_{\nu_{\tau}}(10^{-2}eV)$ \\
  \hline
  0.75 & 1.297 & 2.487 & -3.677 & 3.784 & 6.271 \\
  0.73 & 1.271 & 2.637 & -4.003 & 3.908 & 6.346 \\
  0.71 & 1.245 & 2.789 & -4.333 & 4.034 & 6.424 \\
  0.69 & 1.220 & 2.943 & -4.666 & 4.163 & 6.506 \\
  \hline
\end{tabular}}
\end{table}

Without losing generality, considering the case $v_1^{\nu} \ll
v_2^{\nu}, v_3^{\nu}$, we then have the following approximate
relations
\begin{eqnarray}
& & \Delta \simeq \frac{3r}{(1+r)^2},\quad \tan 2 \theta_{\nu}
\simeq \frac{\sqrt{3}(1-r)}{1+r}, \quad r \equiv v_3^{\nu}/v_2^{\nu}
\end{eqnarray}
which shows that in this case both the mixing angle $\theta_{\nu}$
and the ratio $r$ can be determined for the given values of
$\Delta$, an interesting solution is
\begin{eqnarray}
\Delta = r = \sqrt{3} - 1, \quad \tan 2\theta_{\nu} = 2-\sqrt{3},
\quad \theta_{\nu} = 7.5^{\circ}
\end{eqnarray}
For a numerical estimation, it is useful to investigate the
following two interesting cases
\begin{eqnarray}
& & \mbox{Case I}:\qquad s_{13}^e\simeq 0,\ s_{12}^e\simeq 0 \\
& & \mbox{Case II}:\qquad s_{13}^e \ll s_{12}^e \sim
\sqrt{m_e/m_{\mu}} \simeq 0.07, \quad \delta_1^e - \delta_2^e =\pi/2
\end{eqnarray}
which allows us to present a reasonable estimation for $\theta_{13}$
and $\delta_{\nu}$ (see table 2).

\begin{table}[ph]
\tbl{The mixing angle $\theta_{13}$ and CP phase $\delta_{nu}$ as
function of vacuum parameter $\Delta$}
{\begin{tabular}{@{}|c|c|c|c|c|c|c|@{}}
  \hline
  $\Delta$ (input) & $r=v_3^{\nu}/v_2^{\nu}$ & $\theta_{\nu}$  & $\ \theta_{13}$ (Case I) & $\ \delta_{\nu}$ (Case I)
   & $\ \theta_{13}$ (Case II) & $\ \delta_{\nu}$ (Case II) \\
  \hline
  0.75 & 1.0 & 0 & 0 & 0 & $2.8^{\circ}$ & $90^{\circ}$ \\
  0.748 & 0.90 & $2.6^{\circ}$ & $2.1^{\circ}$ & 0 & $3.5^{\circ}$ & $54^{\circ}$ \\
  0.745 & 0.85 & $4.0^{\circ}$ & $3.3^{\circ}$ & 0 & $4.4^{\circ}$ & $41^{\circ}$ \\
  0.74 & 0.79 & $5.8^{\circ}$ & $4.7^{\circ}$ & 0 & $5.5^{\circ}$ & $31^{\circ}$ \\
  0.73 & 0.72 & $7.9^{\circ}$ & $6.4^{\circ}$ & 0 & $7.0^{\circ}$ & $24^{\circ}$ \\
  0.71 & 0.63 & $10.7^{\circ}$ & $8.7^{\circ}$ & 0 & $9.2^{\circ}$ & $18^{\circ}$ \\
  0.69 & 0.56 & $13.0^{\circ}$ & $10.6^{\circ}$ & 0 & $11.0^{\circ}$ & $15^{\circ}$ \\
  \hline
\end{tabular} }
\end{table}
where $v_3^{\nu}\neq v_2^{\nu}$ (i.e., $r\neq 1$) should be a more
general and reasonable case when no symmetry is imposed, the
resulting mixing angle $\theta_{13}$ can be large enough to be
detected. For the case II, both mixing angle $\theta_{13}$ and
CP-violating phase $\delta_{\nu}$ are in general testable by the
future neutrino experiments. For the typical range $\Delta \simeq
0.72\pm 0.028$, we are led to the most optimistic predictions for
the mixing angle $\theta_{13}$ and CP-violating phase $\delta_{\nu}$
\begin{eqnarray}
& & \theta_{13}\simeq 7^{\circ}\pm 4^{\circ},\qquad \delta_{\nu}
\simeq 35^{\circ} \pm 20^{\circ}, \quad  r=v_3^{\nu}/
v_2^{\nu}\simeq 0.73\pm 0.17
\end{eqnarray}
which can be tested in the future experiments.

When taking the Dirac type neutrino masses $m_{\nu}^D$ and $m_N^D$
to be at the order of $(0.1\sim 1.0)$ MeV (i.e., at the same order
of electron mass), and the mass parameter $m_E^D \simeq (15 \sim
25)$\ GeV, we have the lightest vector-like Majorana neutrino masses
and charged lepton mass to be
\begin{eqnarray}
& & m_{N_1} = m_{N_3} \simeq O(250)\ \mbox{GeV} \sim O(25)\
\mbox{TeV} \\
& & m_{E_3} \simeq (127 \sim 352)\ \mbox{GeV}
\end{eqnarray}
which is at the electroweak scale and can be explored at LHC and
ILC.

In conclusion, we have shown how the puzzles on the smallness of
neutrino masses and the nearly tri-bimaximal neutrino mixing may
simultaneously be understood in the flavor SO(3) gauge symmetry
model with the mechanism of approximate global U(1) family symmetry.
The vacuum structure of SO(3) symmetry breaking for the SO(3)
tri-triplet Higgs bosons plays an important role. The mixing angle
$\theta_{13}$ is in general nonzero and its typical values range
from the experimentally allowed sensitivity to the current
experimental bound. CP violation in the lepton sector is caused by a
spontaneous symmetry breaking and can be significantly large for
exploring via a long baseline neutrino experiment. A similar
consideration can be extended to the quark sector, unlike the lepton
sector with the features of Majorana neutrinos, the mechanism of
approximate global U(1) family symmetry can be applied to understand
the smallness of quark mixing angles.

\section*{Acknowledgments}

\label{ACK}

This work was supported in part by the National Science Foundation
of China (NSFC) under the grant 10475105, 10491306, and the key
Project of Chinese Academy of Sciences (CAS). The author is grateful
to Chun Liu for his hard work on publishing the proceedings of
ICFP2007.

%\end{document}


\begin{thebibliography}{99}
\bibitem{YLW0}Y.L. Wu, Phys. Rev. {\bf D}77 113009 (2008),
arXiv:0708.0867.
\bibitem{JN} J. Ng, a summary talk, in this proceedings.
\bibitem{EXP1}B. Aharmim et al. (SNO Collaboration),
Phys.Rev. C72 (2005) 055502.
\bibitem{EXP2} K. Eguci et al. (KamLAND Collaboration),
Phys. Rev. Lett. {\bf 94}, 081801(2005).
\bibitem{EXP3} E.Aliu et al.(K2K Collaboration), Phys. Rev. Lett. {\bf 94}, 081802(2005)
\bibitem{EXP4} Y. Ashie et al. ( SK Collaboration),  Phys.Rev. D71 (2005).
\bibitem{EXP5} M. Altmann et al. (GNO Collaboration), Phys. Lett. {\bf B616}, 174(2005).
\bibitem{EXP6} M. Ambrosio et al. (MARCO Collaboration), Eur. Phys. J. {\bf C36}, 323(2004).
\bibitem{EXP7} M. Sanchez et al.( Soudan 2 Collaboration), Phys. Rev. {\bf D68},
113004(2003).
\bibitem{EXP8} M. Apollonio et al. (CHOOZ Collaboration), Eur. Phys.
J. {\bf C27}, 331(2003).
\bibitem{PDG} W.-M. Yao, et al., Particle Data Group, Journal of Phys.
{\bf G33}, 1 (2006).
\bibitem {SV}A. Strumia and F. Vissani, arXiv:hep-ph/0606054.
\bibitem{TNM2}   G.L. Fogli, E. Lisi, A. Marrone, A. Palazzo, Prog. Part. Nucl. Phys. {\bf 57} (2006) 742-795.
\bibitem{TNM1}M. Maltoni, T. Schwetz, M. A. Tortotla and J. W. F. Valle, Phys. Rev. {\bf D68},
113010 (2003).
\bibitem{HF} H. Fritzsch, in this proceedings.
\bibitem{PM} P. Minkowski, in this proceedings.
\bibitem{MS} For a recent review see: R. N. Mohapatra and A. Y. Smirnov, Ann. Rev. Nucl. Part. Sci. {\bf 56} 569
 (2006).
\bibitem{HPS}P. F. Harrison, D. H. Perkins and W. G. Scott, Phys. Lett.
{\bf B 530}, 167 (2002).
\bibitem{HPS1} Z.-Z. Xing, Phys. Lett. {\bf B533}, 85(2002).
\bibitem{HPS2} P. F. Harrison and W.G. Scott, Phys. Lett. {\bf B535},
163(2002).
\bibitem{HPS3} P.F. Harrison and W.G. Scott, Phys. Lett. {\bf B557},
76(2003).
\bibitem{HPS4} X. G. He and A. Zee, Phys. Lett. {\bf B560}, 87(2003).
%\bibitem{TBM1} C.I. Low and R. R. Volkas, Phys. Rev. {\bf D68}, 033007
%(2003).
%\bibitem{TBM2} E. Ma,  Phys. Rev. {\bf D70}, 031901R(2004);
%\bibitem{TBM3} G. Altarelli and F. Feruglio, Nucl. Phys. {\bf B720}, 64(2005);
%\bibitem{TBM4} E. Ma, Phys. Rev. {\bf D72}, 037301 (2005).
%\bibitem{TBM5} E. Ma,  Mod.\ Phys.\ Lett.\ A {\bf 20}, 2601 (2005).
%\bibitem{TBM6} A. Zee, Phys. Lett. {\bf B630}, 58 (2005).
%\bibitem{TBM7} E. Ma,  Phys.\ Rev.\ D {\bf 73}, 057304 (2006).
%\bibitem{TBM8} G. Altarelli and F. Feruglio, Nucl. Phys. {\bf B741},
%215(2006).
%\bibitem{TBM9} W. Grimus and L. Lavoura, {\bf JHEP}, 0601:018(2006).
%\bibitem{TBM10} J.E. Kim and J.-C. Park, {\bf JHEP} 0605:017(2006).
%\bibitem{TBM11} N. Singh, M. Rajkhowa and A. Borach, hep-ph/0603189.
%\bibitem{TBM12} R. Mohapatra, S. Naris and Y.-H. Yu, Phys.Lett. {\bf B639} 318
%(2006).
%\bibitem{TBM13} P. Kovtun and A. Zee, Phys.Lett. {\bf B640} (2006)
%37.
%\bibitem{TBM14} N. Haba, A. Watanabe and K. Yoshioka, Phys.Rev.Lett. 97 (2006)
%041601.
%\bibitem{TBM15} X.G. He, Y.Y. Keum and R. Volkas, {\bf JHEP},
%0604:039(2006).
%\bibitem{TBM16} I. Varizelas, S.-F. King and G.G. Ross, Phys.Lett.
%B644 (2007) 153.
%\bibitem{TBM17} E.Ma, hep-ph/0701016 .
\bibitem{Ma} E. Ma, in this proceedings.
\bibitem{Lam} C.S. Lam, in this proceedings.
\bibitem{YLW1} Y.L. Wu,  Phys.Rev. {\bf D60} (1999) 073010.
\bibitem{YLW2} Y.L. Wu, Nucl.Phys.Proc.Suppl. {\bf 85} (2000)
193.
\bibitem{YLW3}Y.L. Wu, invited talk at the 30th International Conference on
High-Energy Physics (ICHEP 2000), Osaka, Japan.
\bibitem{YLW4} Y.L. Wu, Eur.Phys.J. {\bf C10} (1999) 491.
\bibitem{YLW5} Y.L. Wu, J. Phys. G: Nucl. Part. Phys. {\bf 26} 1131 (2000).
\bibitem{YLW6} Y.L. Wu, Science in China {\bf A43} (2000)
988.
\bibitem{CS} C. Carone and M. Sher, Phys. Lett. {\bf B420}, 83 (1998).
\bibitem{MA} E. Ma, hep-ph/9812344.
\bibitem{CW} C. Wetterich, hep-ph/9812426.
\bibitem{BHKR} R. Barbieri, L.J. Hall, G.L. Kane and G.G. Ross, hep-ph/9901228.
\bibitem{FLee}R. Friedberg and T. D. Lee, arXiv:hep-ph/0606071;
arXiv:hep-ph/0705.4156.
\bibitem{HWW} B.Hu, F. Wu and Y.L. Wu, Phys.Rev. {\bf D75} 113003
(2007).
\bibitem{HW} L.J. Hall and S. Weinberg, Phys. Rev. {\bf D48}, 979 (1993).
\bibitem{WW} Y.L. Wu and L. Wolfenstein, Phys. Rev. Lett. {\bf 73} 1762 (1994).
\bibitem{WU} Y.L. Wu, Carnegie-Mellon Univ. report, CMU-HEP94-01, hep-ph/9404241, 1994;
\\ Invited talk at 5th Conference on the Intersections
of Particle and Nuclear Physics, St. Petersburg, FL, 31 May- 6 Jun
1994, published in Proceedings, pp338, edited by S.J. Seestrom (AIP,
New York, 1995), hep-ph/9406306.
\end{thebibliography}
\end{document}